\documentclass{aipproc}

\layoutstyle{8x11double}

\title{The Role of Dust in GRB Afterglows}

\author{Donald Q. Lamb}
{
address={Department of Astronomy \& Astrophysics, University of Chicago,
5640 South Ellis Avenue, Chicago, IL 60637},
email={lamb@oddjob.uchicago.edu},
}

\begin{abstract}
We show that the clumpy structure of star-forming regions can naturally
explain the fact that 50-70\% of GRB afterglows are optically ``dark.'' 
We also show that dust echos from the GRB and its afterglow, produced
by the clumpy structure of the star-forming region in which the GRB
occurs, can lead to temporal variability and peaks in the NIR, optical,
and UV lightcurves of GRB afterglows.  We note that the detection
of GRB ``orphan'' afterglows would provide strong evidence that the
star-forming regions in which GRBs occur are clumpy.
\end{abstract}

\begin{document}

\maketitle

\section{Introduction} 

There is increasing evidence that the long gamma-ray bursts (GRBs) are
associated with galaxies undergoing copious star formation, and occur
near or in the star-forming regions of these galaxies (see, e.g., [1]
for a discussion of this evidence).  Star-forming regions contain large
amounts of dust that can extinguish the optical and UV light of GRB
afterglows.  Indeed, no optical afterglows have been detected for
60-70\% of the long GRBs.  Some of these failures may be due to the
relatively large size of the GRB error box, or to a delay in observing
the error box.  Some may be because the GRB lies at a very high
redshift, and the Lyman limit lies longward of the optical band [2,3]. 
However, the majority of the failures are most likely because the
optical afterglow is faint or absent due to its extinction by dust in
the host galaxy of the GRB [4]. 

We show that the clumpy structure of star-forming regions can naturally
explain the statistics of optically ``dark'' GRB afterglows.  We also
show that dust echos from the GRB and its afterglow, produced by the
clumpy structure of the star-forming region in which the GRB occurs, 
can lead to temporal variability and peaks in the NIR, optical, and UV 
lightcurves of GRB afterglows.

\begin{figure}[ht]
\includegraphics[scale=0.35,angle=90]{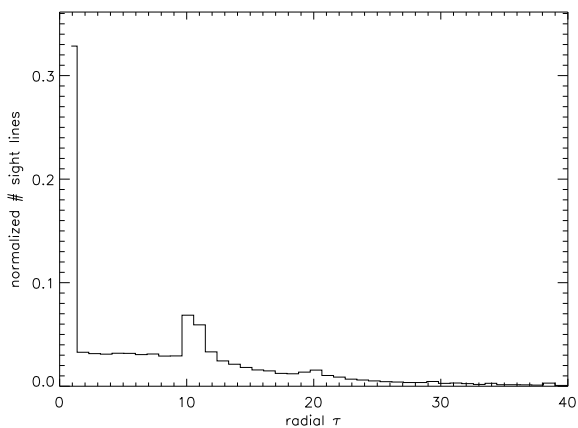}
\caption{Distribution of optical depths, as experienced by random
photons emitted isotropically by a central source in a clumpy medium
with an amount of dust equivalent to a radial optical depth $\tau_H =
10$ in a homogeneous uniform density medium, a volume filling factor $f
= 0.10$ of the clumps, and a density contrast $k_1/k_2 = 100$ between
the clumps and the interclump medium.  From [4].}
\end{figure}

\begin{figure}[ht]
\includegraphics[scale=0.65,clip=]{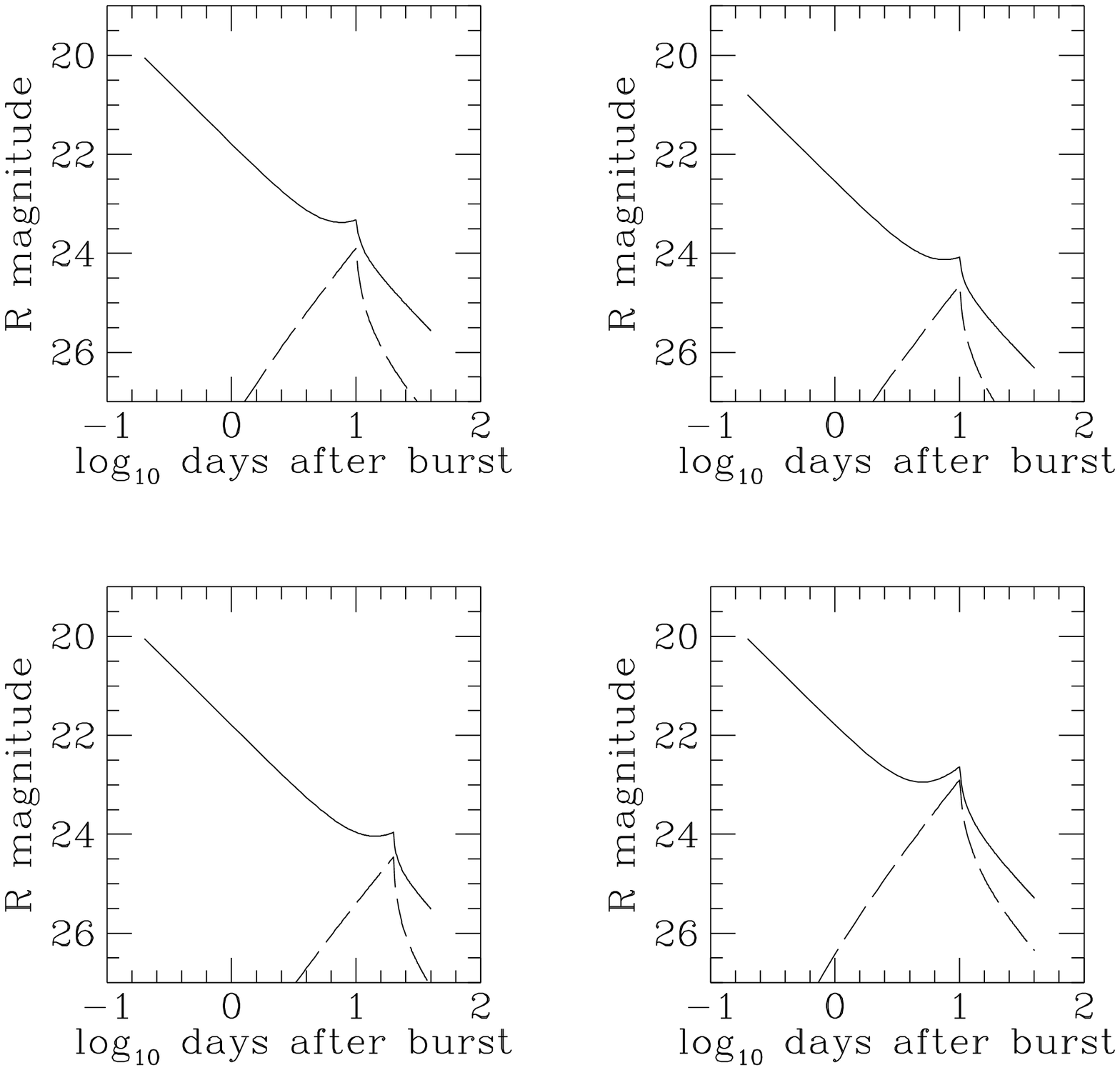}
\caption{Contribution to the observed R-magnitude from the light
forward-scattered by the dust cloud (dashed line), and the sum of this
light and the light seen directly from the point-like source (solid
curve) for two sets of parameters of the simple model
described in the text.  
Left: $\delta t = 20$ days $A = 0.2$, $\epsilon = 0.1$ days.
Right: $\delta t = 10$ days $A = 0.5$, $\epsilon = 0.1$ days.
}
\end{figure}

\section{Structure of Star-Forming Regions}

Star-forming regions are thought to be clumpy, with dense dust clouds
embedded in a much less dense intercloud medium.  A simple model of 
star-forming regions can therefore be characterized by three
parameters: (1) the amount of dust equivalent to a radial optical depth
$\tau_H$ of a homogeneous uniform density medium, the volume filling
factor $f$ of the dust clumps, and the density contrast $k_1/k_2$ of 
the clumps relative to the interclump medium [5,6].  In this model, the
dense dust clumps form connected structures, as is true in real
star-forming regions.

Figure 1 shows the of a Monte Carlo calculation of the distribution of
optical depths for random lines of sight (LOS) from the center of a
clumpy star-forming region to an external observer.  The parameters of
this particular calculation are $\tau_H = 10$, $f = 0.10$, $k_1/k_2 =
100$, and the number $N^3$ of spatial bins is $20^3$ [5].  These 
parameter values lead to results that are consistent with observations 
of star-forming regions in the Milky Way.  Figure 1 shows that 35\% of 
LOS have optical depths $\tau_{\rm obs} = 1$ (the minimum value
possible in  this particular model), while the remaining 65\% of LOS
have $\tau_{\rm  obs} \gg 1$.  The distribution of optical depths can
be understood as  follows.  A substantial fraction of the photons
emitted by the central  source do not encounter a dense clump; these
photons correspond to the LOS with $\tau_{\rm obs} = 1$.  The
remaining photons emitted by the central source encounter one or more
clumps and experience $\tau_{\rm obs} \gg 1$.  The distribution of
$\tau_{\rm obs}$ in this model is consistent with the statistics of
GRB optical afterglows, 50-70\% of which are optically ``dark.''  It
also implies that GRB ``orphan'' afterglows can exist, whereas models
in which the star-forming region is approximately uniform do not.

\section{Dust Echos}

Dust echos from the clumpy structure of the star-forming region can
also produce variability and peaks $1^{\rm d} - 30^{\rm d}$ after the
GRB in the NIR, optical, and UV lightcurves of GRB afterglows.  As an
illustrative example, we have calculated the optical light curve
produced by the dust echo from a single clump at a substantial distance
from the burst source but along the LOS from the burst source to the
observer.  The delay time $\delta t$ between the GRB and the echo is
characterized by $\delta t = (1 + z) R^2_{\perp, {\rm min}}/cD$, where
$D$ is the distance between the GRB and the dust cloud and $R_{\perp,
{\rm min}}$ is the minimum of the perpendicular extent of the dust
cloud $R_{\perp, {\rm cloud}}$, $\theta_{\rm jet}(t) D$, and
$\theta_{\rm forward} D$.  Here $\theta_{\rm jet}(t)$ is the half
opening angle of the afterglow jet and $\theta_{\rm forward}$ is the
half angular width ($\approx 10^\circ-20^\circ$) of the forward
scattering peak for scattering by dust grains.

The amplitude $A$ of the dust echo relative to the direct light from
the afterglow is a function of the albedo $a$ of the dust, the
scattering phase function $\Phi(\theta)$, and the optical depth
$\tau_{\rm clump}$ of the dust clump, all of which are wavelength
dependent.  For forward scattering, which we assume here, $\Phi(\theta)
\approx 1$ [7].  Then in the limiting cases of small and large optical
depths, $A \approx a \tau_{\rm clump}$ and $A \approx a$, respectively.
The prominence of the dust echo also depends on the time $\epsilon$ 
after the GRB that the afterglow begins.

Figure 2 shows the dust echo from a single clump along the LOS from the
burst source to the observer, assuming $\tau_{\rm clump} < 1$.  In the 
two cases shown, the rate of temporal decline of the GRB afterglow was 
taken to be a power law with $b = 1$.  Most GRB afterglows decline
more rapidly with time (i.e., $b = 1.3 - 2.25$).  The dust echo is
more prominent for afterglows that decline more rapidly, since the
contrast between the direct light from the afterglow and the echo --
which reflects the brightness of the afterglow at an earlier time --
is then larger.  Thus the examples we have shown are conservative.  

Studies of the temporal variability of the NIR, optical, and UV
lightcurves of the afterglows of GRBs may allow ``reverberation
mapping'' of the structure of the star-forming regions in which GRBs
occur. 

%Since the echo from the dust clump is due to forward
%scattering, it will have a similar color or will be bluer than the
%direct light from the burst afterglow.  Thus such an echo might account
%for the sharp peak in the optical light curve of the afterglow of GRB
%000103c, but not the peaks in the optical light curves of GRBs 990228
%and 980326 (see also [7]).

\end{document}